\documentclass{article}
\usepackage{spconf,amsmath,graphicx}
\usepackage{math-common}
\usepackage{algorithm}
\usepackage[noend]{algpseudocode}
\usepackage{amsthm}
\usepackage[hidelinks]{hyperref}
\newtheorem*{theorem*}{Theorem}

\newtheoremstyle{cited}%
  {3pt}
  {3pt}
  {\itshape}
  {}
  {\bfseries}
  {.}
  {.5em}
  {\thmname{#1} \thmnumber{#2} \thmnote{\normalfont#3}}

\theoremstyle{cited}
\newtheorem*{citedthm}{Theorem}

\title{AN EFFICIENT ALGORITHM FOR CLUSTERED MULTI-TASK compressive SENSING}
\name{Alexander Lin \qquad Demba Ba\thanks{This work was supported by a NDSEG fellowship and NSF Cooperative Agreement PHY-2019786.  The code is at \url{https://github.com/al5250/multics}.}}
\address{School of Engineering \& Applied Sciences, Harvard University, Boston, MA, USA }
\begin{document}
%
\maketitle
\begin{abstract}
This paper considers clustered multi-task compressive sensing, a hierarchical model that solves multiple compressive sensing tasks by finding clusters of tasks that leverage shared information to mutually improve signal reconstruction.  The existing inference algorithm for this model is computationally expensive and does not scale well in high dimensions.  The main bottleneck involves repeated matrix inversion and log-determinant computation for multiple large covariance matrices.  We propose a new algorithm that substantially accelerates model inference by avoiding the need to explicitly compute these covariance matrices.  Our approach combines Monte Carlo sampling with iterative linear solvers.  Our experiments reveal that compared to the existing baseline, our algorithm can be up to thousands of times faster and an order of magnitude more memory-efficient. 
\end{abstract}

\section{Introduction}
\label{sec:intro}
\emph{Compressive sensing} (CS) is a fundamental problem with applications in many areas of signal processing, such as medical imaging \cite{lustig2008compressed}, astronomy \cite{wiaux2009compressed}, microscopy \cite{studer2012compressive}, and photography \cite{duarte2008single}.  Given measurements $\bd y \in \R^N$ and a sensing matrix $\mathbf \Phi \in \R^{N \times D}$ with $N < D$, the goal is to find a sparse signal $\bd z \in \R^D$ that satisfies $\bd y = \mathbf \Phi \bd z + \text{noise}$.
Certain applications (e.g. multi-contrast MRI \cite{bilgic2011multi}) encounter multiple CS problems $(\bidx y 1, \bidx \Phi 1), \ldots, (\bidx y T, \bidx \Phi T)$, which require corresponding solutions $\bidx z 1, \ldots, \bidx z T \in \R^D$.  \emph{Multi-task compressive sensing} \cite{ji2008multitask} is a popular approach for settings in which $\{\bidx z t\}_{t=1}^T$ are known to have the same non-zero support.  This model leverages shared information to jointly solve the $T$ tasks, outperforming methods that solve each task separately.

In many situations, it is unreasonable to assume that all tasks have the same sparsity structure.  Using multi-task compressive sensing for unrelated tasks with differing supports among $\{\bidx z t\}_{t=1}^T$ can lead to worse performance than solving each task separately \cite{qi2008multi}. Thus, given multiple CS problems, we would like to ideally learn the structure of task relationships and only share information between related tasks; this allows for improvement in overall performance for all tasks.

We consider \emph{clustered multi-task compressive sensing}, a hierarchical model that solves multiple CS problems in which the inter-relatedness of the solutions are apriori unknown.  The model automatically determines how to divide the $T$ tasks into $C$ clusters, only sharing information between related tasks within a cluster.  Although variations of this model have been studied before \cite{qi2008multi, wang2015novel}, the standard inference algorithm is too computationally demanding, making it impractical for high dimensions $D$ and large number of tasks $T$.  The main bottleneck is repeated matrix inversion and log-determinant calculation for many $D \times D$ covariance matrices, requiring $O(T CD^2)$-space and $O(TCD^3)$-time per iteration.  

We propose a new algorithm that is more efficient, reducing the space complexity to $O(TCD)$ and the time complexity to $O(TC \tau_D)$, where $\tau_D \leq O(D^2)$ is the time needed to multiply a vector by the sensing matrix $\bold \Phi$.  The key idea is to use techniques from Monte Carlo sampling and numerical linear algebra to circumvent the need to form the large covariance matrices.  
Our algorithm extends and generalizes the state-of-the-art method that we previously developed for Bayesian compressive sensing \cite{lin2022covariance} to a more challenging setting involving mixture models \cite{mclachlan1988mixture}.  In experiments, we show that our algorithm can be up to thousands of times faster than EM, reducing hours of computation to a few seconds.

\section{MODEL AND BACKGROUND}
\label{sec:format}
The \emph{clustered multi-task compressive sensing} model for $T$ tasks and $C$ clusters has the following generative structure: 
\begin{align}
\idx a t &\sim \mathrm{Categorical}(\idx \pi 1, \ldots, \idx \pi C), \label{cmtcs} \\
\bidx z t \given \idx a t = c, \bidx \alpha c &\sim \mathcal{N}(\bd 0, \mathrm{diag}(\bidx \alpha c)^{-1}), \nonumber \\
\bidx y t \given \bidx z t &\sim \mathcal{N}(\bidx \Phi t \bidx z t, \tfrac{1}{\beta} \bold I),  \quad  t = 1, \ldots, T. \nonumber
\end{align}
This model posits that each CS task $t$ has an unknown cluster assignment $\idx a t$ that is one of $C$ options.  The $C$ clusters have prior probabilities $\idx \pi 1, \ldots, \idx \pi C \in [0, 1]$, which sum to one.  For simplicity, we assume that $\{\idx \pi k\}_{k=1}^K$ are known, but they can also be learned by the model \cite{mclachlan1988mixture}.  All tasks within a particular cluster $c$ are interrelated and share a common non-negative \emph{regularization parameter} $\bidx \alpha c \in \R^D$, which the model will learn through an inference algorithm.  For a particular dimension $d$, if $\idx[d] \alpha c$ is large, then latent vectors $\bidx z t$ within cluster $c$ have small variance $1 / \idx[d] \alpha c$ around zero for that dimension; in the limiting case where $\idx[d] \alpha c \to \infty$, we have exact sparsity, i.e. $\idx[d] z t \to 0$ for all $t$ such that $\idx a t = c$.  Finally, given each $\bidx z t$, the model lets the measurements be $\bidx y t = \bidx \Phi t \bidx z t + \bidx \varepsilon t$ for Gaussian noise $\bidx \varepsilon t \sim \mathcal{N}(0, 1 / \beta)$. 

Eq. \eqref{cmtcs} is a combination of Bayesian compressive sensing \cite{ji2008bayesian} and mixture modeling \cite{mclachlan1988mixture}. For $C = 1$, it reduces to the well-known multi-task CS model of \cite{ji2008multitask}, in which every task has the same parameter.  However, when $C > 1$, it allows for the more general case in which not all $T$ tasks have the same sparsity structure.  
 This model can be fit to data $\{\bidx y t\}_{t=1}^T$ by maximizing the log-likelihood $L(\bd \theta) := 1/T \sum_{t=1}^T \log p(\bidx y t \given \bd \theta)$ of the parameters $\bd \theta := \{\bidx \alpha c\}_{c=1}^C$.  The objective $L$ is known to encourage some components of each $\bidx \alpha c$ to diverge to $\infty$, sparsifying $\bidx z 1, \ldots, \bidx z T \given \bd \theta$ \cite{yee2017sparse}.    

The standard algorithm for optimizing $L$ is \emph{expectation-maximization} (EM) \cite{dempster1977maximum, wang2015novel}, which cycles between two steps.  Given estimates $\bd {\widehat \theta} := \{\bidx {\widehat \alpha} c\}_{c=1}^C$ from the previous iteration, the E-step creates a surrogate $Q(\bd \theta | \bd  {\widehat \theta})$ that lowerbounds $L(\bd \theta)$.  
The M-step maximizes this surrogate to find new parameters $\bd {\widetilde \theta} := \arg \max_{\bd \theta} Q(\bd \theta | \bd {\widehat \theta})$.  
EM theory guarantees that  $L(\bd {\widetilde \theta}) \geq L(\bd {\widehat \theta})$ \cite{dempster1977maximum}, which means that we can optimize $L$ by repeating these two steps.  For the model in Eq. \eqref{cmtcs}, the $Q$-function is
\begin{align}
&Q(\bd \theta | \bd {\widehat \theta}) \hspace{-0.2em} := \hspace{-0.2em} \frac{1}{T}\sum_{t=1}^T \mathbb{E}_{\idx a t , \bidx z t |\bidx y t, \bd {\widehat \theta}}[\log p(\idx a t , \bidx z t, \bidx y t | \bd {\theta})] \hspace{-0.8em} \label{q-func} \\
 & \cong \frac{1}{TC} \sum_{\substack{t=1, \\ c = 1}}^{T, C} q^{(t, c)} \hspace{-0.3em}  
\left[\sum_{d=1}^D \frac{\log \idx[d] \alpha c}{2} - \frac{\idx[d] \alpha c}{2}  \hspace{-0.2em} \left[ (\idx[d] \mu {t, c})^2 
 + \idx[d, d] \Sigma {t, c}\right] \right] \nonumber
\end{align}
where $q^{(t, c)} := p(\idx a t = c \given \bidx y t, \bd {\widehat \theta}) \in [0, 1]$ is the posterior probability of assigning task $t$ to cluster $c$; $\bd \mu^{(t, c)} \in \R^D$ and $\bd \Sigma^{(t, c)} \in \R^{D \times D}$ are the mean and covariance of the Gaussian conditional posterior density $p(\bidx z t \given \idx a t = c, \bidx y t, \bd {\widehat \theta})$; and $\cong$ denotes equality up to additive constants with respect to $\bd \theta$.  The analytic forms of $\bidx \Sigma {t, c}, \bidx \mu {t, c}, q^{(t, c)}$ are
\begin{align}
&\bidx \Sigma {t, c} := \left(\beta ({\bidx \Phi {t}})^\top {\bidx \Phi {t}} + \mathrm{diag}(\bidx {\widehat \alpha} c)\right)^{-1}, \quad \forall t,c \label{estep} \\
&\bidx \mu {t, c} := \beta \bidx \Sigma {t, c} ({\bidx \Phi {t}})^\top \bidx y t, \quad \forall t, c\nonumber \\
& \hspace{-0.4em} \begin{bmatrix}
\idx q {t, 1} \\
\vdots \\
\idx q {t, C} 
\end{bmatrix} := \mathrm{softmax}\left(
\begin{bmatrix}
\frac{1}{2} \idx \ell {t, 1} + \log \idx \pi 1  \\
\vdots \\
\frac{1}{2} \idx \ell {t, C}  + \log \idx \pi C
\end{bmatrix} \right), \quad \forall t 
 \nonumber \\
&\idx \ell {t, c} := \log \det \boldidx \Sigma {t, c} + \sum_{d=1}^D \log \idx[d] {\widehat \alpha} c  + \beta (\bidx y t)^\top \bidx \Phi t \bidx \mu {t, c}\nonumber
\end{align}
By differentiating Eq. \eqref{q-func} with respect to $\bd \theta$, we derive the M-step update for new parameters $\bd {\widetilde \theta} := \{\bidx {\widetilde \alpha} c\}_{c=1}^C$ as
\begin{align}
\idx[d] {\widetilde \alpha} c := \frac{\sum_{t=1}^T q^{(t, c)}}{\sum_{t=1}^T q^{(t, c)} \cdot [(\idx[d] \mu {t, c})^2 + \idx[d, d] \Sigma {t, c}]}, \quad \forall c, d. \label{mstep}
\end{align}
EM repeats Eq. \eqref{estep} and \eqref{mstep} until convergence. Eq. \eqref{estep} is expensive because it involves matrix inversion and log-determinant calculation for the large covariance matrix $\bold \Sigma^{(t, c)}$, which requires $O(D^3)$-time and $O(D^2)$-space.  For $T$ tasks and $C$ clusters, $TC$ different covariances must be computed.  Thus, EM does not scale well for large $D$.  In the next section, we introduce a new method that makes EM more efficient.

\section{Algorithm}
Our proposed algorithm is called \emph{covariance-free expectation-maximization} (CoFEM). It accelerates EM by avoiding explicit formation of the large covariance matrices $\bidx \Sigma {t, c}$.  This algorithm extends our previously developed method for single-task CS \cite{lin2022covariance, lin2022high} to the more complicated multi-task mixture model of Eq. \eqref{cmtcs}.  Our key observation is that Eq. \eqref{mstep} only  requires three types of quantities that depend on  $\bidx \Sigma {t, c}$: the posterior means $\bidx \mu {t, c}$, the posterior variances (i.e. diagonal elements of $\bidx \Sigma {t, c}$), and the posterior cluster probabilities $\idx q {t, c}$.  We will show how to estimate these quantities while avoiding the $TC$ matrix inversions and log-determinant calculations of standard EM.  Our approach combines advances in Monte Carlo sampling and numerical linear algebra.  

\subsection{Estimating Posterior Means and Variances} \label{sec:mean-var}
We begin by describing how to estimate $\bd \mu$ and the diagonal elements of $\bd \Sigma$, following \cite{lin2022covariance} (for notational simplicity we omit the superscripts $(t)$ and $(c)$ in this section, though they are assumed).  First, let $\bidx p 1, \ldots, \bidx p K \in \{+1, -1\}^D$ be $K$ random probe vectors, each containing $D$ independently-drawn Rademacher entries (i.e. $+1$ or $-1$ with equal probability).  Next, following the \emph{diagonal estimation rule} \cite{bekas2007estimator}, we define 
\begin{align}
\bidx x k := \bd \Sigma \bidx p k,  \forall k, \quad \quad \bd s := \frac{1}{K} \sum_{k=1}^K \bidx p k \odot \bidx x k, \label{diag-est}
\end{align}
where $\odot$ denotes element-wise product.  It follows that $\bd s$ is an unbiased Monte Carlo estimator of the diagonal of $\bold \Sigma$ (i.e. $\E[s_d] = \Sigma_{d, d}$ for all $d$) \cite{bekas2007estimator}.  Observe that $\bd \mu$ (Eq. \eqref{estep}) and $\{\bidx x k\}_{k=1}^K$ (Eq. \eqref{diag-est}) are the results of multiplying vectors by $\bd \Sigma$; equivalently, they are the solutions to the linear systems
 \begin{align}
 \bold A \bidx x k = \bidx p k, \forall k, \quad \quad \bold A \bd \mu = \beta \bd \Phi^\top \bd y,   \label{lin-sys}
 \end{align}
 for $\bold A := \bold \Sigma^{-1} = \beta \bold \Phi^\top \bold \Phi + \text{diag}(\bd {\widehat \alpha})$.  We use $U$ steps of the \emph{conjugate gradient (CG) algorithm} (Alg. \ref{conj-grad}) \cite{hestenes1952methods} to solve these systems without forming the matrix $\bold \Sigma$.     
 Thus, CoFEM efficiently estimates $\bd \mu$ and the diagonal elements of $\bold \Sigma$ while avoiding costly matrix inversions. As shown  by \cite{lin2022covariance}, typically small values of $K, U \ll D$ suffice for accurate estimation.   

\setlength{\textfloatsep}{1em}

\begin{algorithm}
\caption{Conjugate Gradient to Solve $\bold A \bd x = \bd b$} \label{conj-grad}
\begin{algorithmic}[1]
\State{Let $\bd x_1 := \bd 0$, $\bd r_1 := \bd b - \bold A \bd x_1$, $\bd d_1 := \bd r_1$.}
\For{step $u = 1, 2, \ldots, \text{until convergence}$}
    \State{$\gamma_u := (\bd r_u^\top \bd r_u) / (\bd d_u^\top \bold A \bd d_u)$}
    \State{$\bd x_{u+1} := \bd x_u + \gamma_u  \bd d_u$}
    \State{$\bd r_{u+1} := \bd r_u - \gamma_u  \bold A \bd d_u$}
    \State{$\xi_{u} := (\bd r_{u+1}^\top \bd r_{u+1}) / (\bd r_{u}^\top \bd r_{u})$}
    \State{$\bd d_{u+1} := \bd r_{u+1} + \xi_u \bd d_u$}
\EndFor
\end{algorithmic}
\end{algorithm}

\begin{algorithm}[h]
\caption{CoFEM for Clustered Multi-Task CS} \label{cofem}
\begin{algorithmic}[1]
\State{Initialize $\{\bidx \alpha c\}_{c=1}^C$ from $\mathrm{Uniform}(0, 1)$.}
\For{\text{each EM iteration until convergence}}
    \For{task $t = 1, \ldots, T$ and cluster $c = 1, \ldots, C$}
        \State{Define $\boldidx A {t, c} \gets \beta (\boldidx \Phi t)^\top\boldidx \Phi t + \mathrm{diag}(\bidx \alpha c)$.}
        \State{Sample Rademacher probes $\bidx p {t, c, 1}, \ldots, \bidx p {t, c, K}$.}
        \State{Solve for $\bidx \mu {t, c}, \{\bidx x {t, c, k}\}_{k=1}^K$ (Eq. \eqref{lin-sys}) via CG.}
        \State{Compute $\bidx s {t, c} \gets 1 / K \sum_{k=1}^K \bidx p {t, c, k} \odot \bidx x {t, c, k}$.}
        \State{Build and eigendecompose $\{\boldidx {\bar{\Gamma}} k\}_{k=1}^K$ (Eq. \eqref{tridiag}).}
        \State{Compute $\bar{\nu}(\boldidx \Sigma {t, c}) \approx \log \det (\boldidx \Sigma {t, c})$ (Eq. \eqref{logdet-est}).}
        \State{Compute $\idx q {t, c}$ from $\bar{\nu}(\boldidx \Sigma {t, c}), \bidx \mu {t, c}$ (Eq. \eqref{estep}).}
    \EndFor{}
    \State{Update $\{\bidx \alpha c\}$ using
    $\{\bidx \mu {t, c}, \bidx s {t, c}, \idx q {t, c}\}$} (Eq. \eqref{mstep}).
\EndFor
\For{task $t = 1, \ldots, T$}
    \State{Find cluster assignment $a_t \gets \arg \max_{c=1}^C \idx q {t, c}$.}
\EndFor \\
\Return reconstructions $\{\bidx \mu {t, a_t}\}_{t=1}^T$
\end{algorithmic}
\end{algorithm}

\subsection{Estimating Posterior Cluster Probabilities}
Unlike the single-task setting of \cite{lin2022covariance},  clustered multi-task CS  also requires posterior cluster probabilities $\idx q {t, c}$.  As shown by Eq. \eqref{estep}, most of the quantities in the definition of  $\idx q {t, c}$ are straightforward to compute; the main remaining bottleneck is the $\log \det \bidx \Sigma {t, c}$ term.  Directly computing a log-determinant is an expensive operation costing $O(D^3)$-time.   Instead, we show how to leverage the probe vectors and CG outputs from diagonal estimation (Eq. \eqref{lin-sys}) to efficiently estimate $\log \det \bidx \Sigma {t, c}$ in $O(KU^3)$-time, where $K, U \ll D$.  Let $\bold A := \bold \Sigma^{-1}$ and $\bold L := \log \bold A$ be its matrix logarithm.  Then, we have $\log \det \bold \Sigma = -\log \det \bold A = -\mathrm{tr}(\bold L)$ \cite[Th. 2.12]{hall2013lie}.  We propose to estimate $-\mathrm{tr}(\bold L)$ using the \emph{Hutchinson trace estimator} $\nu := -1 / K  \sum_{q=1}^K (\bidx p k)^\top \bold L \bidx p k$, where $\{\bidx p k\}_{k=1}^K$ are Rademacher probe vectors \cite{hutchinson1989stochastic}.  This estimator is unbiased, meaning that $\mathbb{E}[\nu] = -\mathrm{tr}(\bold L) = \log \det \bold \Sigma$. 

The final question is how to obtain $\bd p^\top \bold L \bd p$, where $\bd p$ is a Rademacher probe (we drop the index $(k)$ for simplicity).  We adapt the methods of Sec. 3.1 to compute a \emph{Lanczos quadrature} \cite{ubaru2017fast, gardner2018gpytorch}:  Recall that in Eq. \eqref{lin-sys}, we use CG to solve the system $\bold A \bd x = \bd p$.  CG theory shows that $D$ steps of Alg. \ref{conj-grad} implicitly performs a Lancozs tridiagonalization $\bold A = \bold R^\top \bold \Gamma \bold R$ \cite[Sec. 6.7.3]{saad2003iterative}, where $\bold R \in \R^{D \times D}$ is an orthonormal matrix whose rows are normalized CG residuals $\{\bd r_u / \norm{\bd r_u}\}_{u=1}^D$ (line 5) and $\bold \Gamma \in \R^{D \times D}$ is a symmetric tridiagonal matrix assembled from CG step sizes $\{\gamma_u, \xi_u\}_{u=1}^D$ (lines 3 \& 6), i.e.
\begin{align}
\Gamma_{1, 1} := \tfrac{1}{\gamma_1}, \; \; \Gamma_{u, u} := \tfrac{1}{\gamma_u} + \tfrac{\xi_{u-1}}{\gamma_{u-1}}, \; \; \Gamma_{u, u-1} := \tfrac{\sqrt{\xi_{u -1}}}{\gamma_{u-1}} \label{tridiag}
\end{align}
for $1 < u \leq D$. Let $\bold \Gamma := \bold S^\top \mathrm{diag}(\bd \lambda) \bold S$ be the eigendecomposition of $\bold \Gamma$ for eigenvalues $\bd \lambda \in \R^D$ and eigenvectors stored as rows of $\bold S \in \R^{D \times D}$.  Since $\bold A = (\bold S \bold R)^\top \mathrm{diag}(\bd \lambda) \bold S \bold R $ and $\bold L = \log \bold A$, we have $\bold L = (\bold S \bold R)^\top \mathrm{diag}(\log \bd \lambda) \bold S \bold R $.  Thus,
\begin{align}
\hspace{-0.55em} \bd p^\top \bold L \bd p = (\bold S 
\hspace{-0.4em}{\underbrace{\bold R \bd p}_{\sqrt{D} \cdot \bd e_1}} \hspace{-0.3em})\hspace{-0.1em}^\top \hspace{-0.15em} \mathrm{diag}(\log \bd \lambda) \bold S \hspace{-0.4em} {\underbrace{\bold R \bd p}_{\sqrt{D} \cdot \bd e_1}} \hspace{-0.4em} = D\sum_{u=1}^D S_{u, 1}^2 \hspace{-0.1em} \log \lambda_u \label{hutch-trace}
\end{align}
where $\bd e_1 := [1 \;0 \; \cdots \; 0]^\top$ and $\bold R \bd p = \sqrt{D} \cdot \bd e_1$ because using CG to solve $\bold A \bd x = \bd p$ with initial condition $\bd x_1 = \bd 0$ means that the first scaled residual (i.e. first row of $\bold R$) is $\bd p / \norm{\bd p}$ and all subsequent residuals/rows of $\bold R$ are orthogonal to $\bd p$ \cite{gardner2018gpytorch, saad2003iterative}.  

Though we can use Eq. \eqref{hutch-trace} to compute $\bd p^\top \bold L \bd p$, it requires eigendecomposition of the full $D \times D$ matrix $\bold \Gamma$, which is an expensive $O(D^3)$-time operation.  As shown by \cite{lin2022covariance}, typically $U \ll D$ steps of CG suffice for accurate diagonal estimation.  To save computation, we would like to only use these $U$ steps to estimate $\bd p^\top \bold L \bd p$, yet this means that only $\bold {\bar \Gamma} \in \R^{U \times U}$, the upper-left submatrix of $\bold \Gamma$, can be generated by Eq. \eqref{tridiag}.  Let $\bold {\bar S}^\top \mathrm{diag}(\bd {\bar \lambda}) \bold {\bar S}$ be the eigendecomposition of $\bold {\bar{\Gamma}}$, for $\bold {\bar S} \in \R^{U \times U}$ and $\bd {\bar \lambda} \in \R^U$.  The extremal values of $\bd \lambda$, which dominate Eq. \eqref{hutch-trace}, are well-approximated by those of $\bd {\bar \lambda}$ \cite{scales1989use}.  This justifies the following estimator for $\log \bold \Sigma$, which we compute using the byproducts of $U$-step CG for the $K$ systems in Eq. \eqref{lin-sys},
\begin{align}
\bar{\nu}(\bold \Sigma) := - \frac{1}{K} \sum_{k=1}^K \sum_{u=1}^U D \cdot (\idx[u, 1]{\bar{S}}{k})^2\cdot  \log \idx[u] {\bar{\lambda}} k \label{logdet-est}
\end{align}
This estimator is exact for $K = \infty$ probes and $U = D$ CG steps, but has some error if $K, U$ are small. 
For a desired error level, \cite{ubaru2017fast} has the following result on how to choose $K, U$. 

\begin{citedthm}[\cite{ubaru2017fast}]
 Let $\kappa$ be the condition number of $\bold \Sigma$.  For $\epsilon, \eta \in [0, 1]$, if $K \geq \frac{24}{\epsilon^2} \log(\frac{2}{\eta})$ and $U \geq \frac{\sqrt{\kappa}}{4} \log(\frac{\mathcal{O}(\kappa)}{\epsilon \sqrt{\kappa}})$, then
$\mathrm{Pr} (|\bar{\nu}(\bold \Sigma) - \log \det \bold \Sigma| \leq \epsilon \cdot |\log \det \bold \Sigma|) > 1 - \eta$.
\end{citedthm}

\subsection{Full Algorithm and Complexity Analysis}
Alg. \ref{cofem} summarizes the full CoFEM algorithm.  The main  cost is solving $K+1$ systems for each task $t$ and cluster $c$.  We can parallelize CG to solve all $TC(K+1)$ systems simultaneously on multi-core hardware, such as GPUs \cite{lin2022covariance, bekas2007estimator}.  In addition, when solving a system with $\bold A := \beta \bold \Phi^\top \bold \Phi + \mathrm{diag}(\bd \alpha)$, the matrix $\bold A$ never needs to be explicitly computed; CG (Alg. \ref{conj-grad}) only requires a method to compute the matrix-vector product $\bold A \bd v$ for any $\bd v \in \R^D$.  Thus, each CG step costs $\tau_D$, the time complexity of applying the sensing matrix $\bold \Phi$ (and its transpose) to a vector $\bd v$.  Though $\tau_D$ is upperbounded by $O(D^2)$, it can be $O(D \log D)$ or $O(D)$ in many signal processing applications in which $\bold \Phi$ has special structure (e.g. Fourier operators, wavelets, convolutions, low-rank structure, sparsity).

\begin{figure*}[h!]
    \centering
    \includegraphics[scale=0.44]{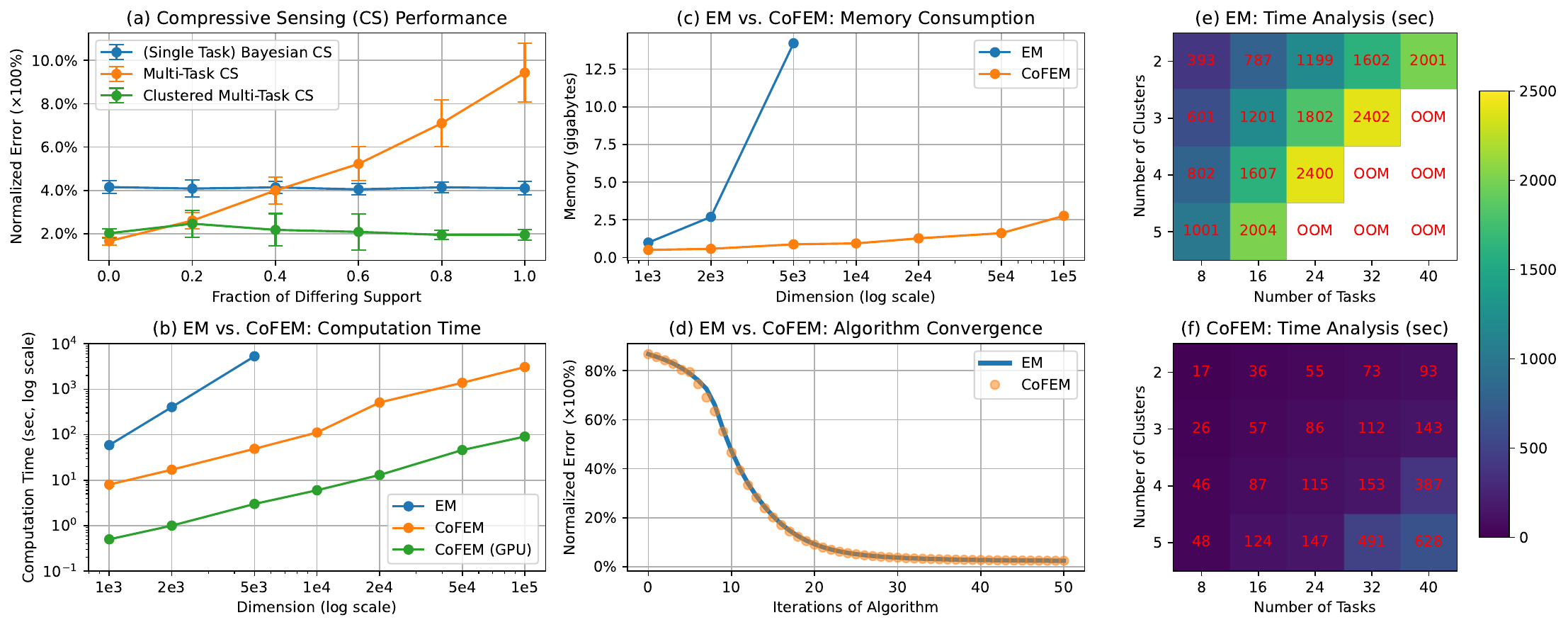}
    \vspace{-0.2em}
    \caption{{(a) Comparison of CS algorithms; (b, c, d, e, f) comparison between EM and CoFEM.  OOM stands for ``out of memory".}}
    \label{fig:main}
    \vspace{-1em}
\end{figure*}

In summary, CoFEM reduces EM's time complexity from $O(TCD^3)$ to $O(TCK(\tau_D U + U^3))$, where for each task, cluster, and probe, $\tau_D U$ is the cost of CG and $U^3$ is the cost of eigendecomposition for the log-determinant.  Additionally, CoFEM reduces EM's space complexity from $O(TCD^2)$ to $O(TCK(D + U))$. As shown in both \cite{lin2022covariance} and Sec. \ref{sec:exp}, $K$ and $U$ can be held as small, constant values even as $D$ becomes very large, allowing CoFEM to scale much better than EM.

\section{Experiments}
\label{sec:exp}

We perform multi-task CS experiments to validate our new algorithm.  For each task $t \in \{1, \ldots, T\}$, we simulate a true sparse signal $\bidx {\widetilde z} t \in \R^D$ with $5\%$ of its components sampled from $\mathcal{N}(0, 1)$ and the rest equal to zero.  Each sensing matrix is $\boldidx \Phi t := \bidx \Omega t \bold F$, where $\bold F \in \mathbb{C}^{D \times D}$ is the discrete Fourier transform and $\boldidx \Omega t \in [0, 1]^{N \times D}$ is an undersampling mask (i.e. randomly-chosen row-wise subset of the identity matrix) with $N = D / 4$.  This form of sensing matrix is common to many CS applications \cite{bilgic2011multi}.  We generate the measurements $\bidx y t = \bidx \Phi t \bidx {\widetilde z} t + \bidx \varepsilon t$, where $\bidx \varepsilon t \sim \mathcal{N}(0, \sigma^2)$ for $\sigma = 0.05$. 

Our first experiment demonstrates the utility of multi-task CS when there exists subsets of interrelated tasks.  We simulate two groups of four tasks (i.e. $T = 8$ total tasks) in which all $\{\bidx {\widetilde z} t\}$ within a group have the same non-zero support set (but still differ in their actual values). 
 For $f \in [0, 1]$, we let the two groups share $1 - f$ of their support and differ in the remaining $f$; thus, if $f = 0$, all tasks have the exact same support and if $f = 1$, there are two groups of tasks with disjoint supports.  We compare how well three methods reconstruct $\{\bidx {\widetilde z} t\}_{t=1}^T$ from the measurements $\{\bidx y t\}_{t=1}^T$: (1) single-task Bayesian CS (which uses a different $\bd \alpha$ per task) \cite{ji2008bayesian}, multi-task CS (which shares a single $\bd \alpha$ across all tasks) \cite{ji2008multitask}, and (3) clustered multi-task CS (which learns $K = 2$ clusters of tasks with cluster-level $\bd \alpha$'s and $\pi_k=1/K$).  We measure success through normalized error $\norm{\btilde z - \bd \mu} / \norm{\btilde z}$, where $\bd {\widetilde z}, \bd {\mu} \in \R^{TD}$ are the concatentations of the true signals $\{\bidx {\widetilde z} t\}_{t=1}^T$ and reconstructions  $\{\bidx \mu t\}_{t=1}^T$.  Fig. 1(a) shows that multi-task CS outperforms single-task CS when all tasks are interrelated (i.e. low $f$), but performs worse when some tasks are unrelated (i.e. high $f$).  In contrast, clustered multi-task CS obtains low error for all $f$, as it automatically learns subsets of related tasks and only shares information among each subset.   

 In the remaining experiments, we show the scalability that CoFEM provides over EM for clustered multi-task CS.  In all cases, we use 50 algorithm iterations, $K = 15$ probe vectors and $U = 50$ CG steps.  Fig. 1(b) and 1(c) show how computation time and memory consumption scale in relation to the problem dimension $D$.  We observe that CoFEM is up to hundreds of times faster and 14 times more memory-efficient than EM. Furthermore, EM cannot be executed for $D > 5{,}000$ due to its high memory usage, yet CoFEM has low memory consumption even for $D = 100{,}000$. Since CoFEM is highly parallelizable, we can also deploy it on parallel hardware (i.e. GPUs).  This brings further acceleration, making CoFEM up to thousands of times faster than EM and reducing hours of computation to a few seconds.   We use a 16-GB, 2.5 GHz Intel Xeon CPU and a Nvidia T4 GPU.    Fig. 1(d) shows that even though CoFEM makes approximations to EM, these approximations are accurate enough to allow for the same rate of convergence in practice.  
 Finally, in Fig. 1(e) and 1(f), we compare how computation time scales in relation to the number of tasks $T$ and the number of clusters $C$ for EM and CoFEM (fixing $D = 2{,}000$).  For larger numbers of tasks and clusters, EM becomes too memory-intensive due to the presence of more covariance matrices.  In contrast, CoFEM experiences no such issues and can process (4 clusters, 40 tasks) in the same time it takes for EM to process (2 clusters, 8 tasks).

\section{Conclusion}

This paper proposed a new algorithm that substantially accelerates inference for multi-task CS mixtures.  Our ideas can be extended to handle infinite mixtures, which place a prior over the number of clusters \cite{wang2015novel}, as well as mixtures of other models (e.g. factor analysis models, time series models) \cite{lin2023probabilistic}.

\newpage
\bibliographystyle{IEEEbib}
\bibliography{refs}

\end{document}